# Growth-controlled photochromism in yttrium oxyhydride thin films deposited by HiPIMS and pulsed-DC magnetron sputtering

M. Zubkins[1,*], E. Letko[1], E. Strods[1], V. Vibornijs[1], D. Moldarev[2], K. Sarakinos[2,3], K. Mizohata[2], K. Kundzins[1], and J. Purans[1]

[1]Institute of Solid State Physics, University of Latvia, Kengaraga 8, LV-1063, Riga, Latvia

[2]Department of Physics, University of Helsinki, P.O. Box 43, FI-00014 Helsinki, Finland

[3]KTH Royal Institute of Technology, Department of Physics, Roslagstullsbacken 21, 11421 Stockholm, Sweden

*Corresponding author: martins.zubkins@cfi.lu.lv

**Abstract**

The present study investigates photochromic oxygen-containing yttrium hydride (YHO) thin films deposited by reactive high power impulse magnetron sputtering (HiPIMS) and compares their photochromic, optical, and structural properties with those of films synthesized by reactive pulsed direct current magnetron sputtering (pulsed-DCMS). Optical emission spectroscopy reveals that, unlike pulsed-DCMS where $Ar^+$ ions dominate, HiPIMS discharges are characterised by strong $Y^+$ emission, evidencing high yttrium ionisation and substantial self-sputter recycling. The critical working pressure ($P_c$) required to obtain transparent and photochromic films is higher for HiPIMS ($P_c \approx 1.0$ Pa) than for pulsed-DCMS ($P_c \approx 0.5$ Pa). Although films deposited near $P_c$ exhibit similar solar transmittance (~72 %) and lattice parameters (5.38–5.39 Å), the pulsed-DCMS film shows a substantially higher relative photochromic contrast (34 %) and a lower optical band gap (2.70 eV) compared with the HiPIMS film (9 % contrast and 2.94 eV). This difference is partly attributed to a lower oxygen-to-hydrogen atomic ratio in the pulsed-DCMS film. Structurally, HiPIMS films are largely polycrystalline with random out-of-plane crystallographic orientation, whereas pulsed-DCMS films exhibit a pronounced <100> out-of-plane preferred orientation. These results demonstrate that, beyond composition, thin-film growth conditions and microstructure play a crucial role in governing the photochromic performance of YHO.

**Introduction**

Buildings account for a substantial share of global energy consumption and greenhouse gas emissions, largely due to heating, cooling, and lighting [1]. Dynamic glazing technologies, or smart windows, offer a promising route to reduce this footprint by regulating solar heat and visible light transmission in response to external conditions [2]. Chromogenic materials, which reversibly modulate their optical properties under external stimuli, are being actively explored for next-generation smart windows. Among these, photochromic thin films respond directly to sunlight without requiring electrical wiring or external control. Oxygen-containing yttrium hydride (hereinafter YHO), also referred to as yttrium oxyhydride, is a material that exhibits a positive photochromic effect under ambient conditions [3]. In the clear state, YHO films typically display an average visible-light transmittance of 70–80% [4]. Upon illumination with a solar simulator, they undergo a colour-neutral darkening with contrast levels of up to 45% [5], followed by gradual self-bleaching once the illumination ceases. The magnitude and kinetics of the photochromic response are strongly dependent on factors such as chemical composition [6], anionic ordering in the sublattice [7], the presence of defects [8,9], film thickness [10], and illumination conditions [11]. Importantly, YHO films can be deposited by magnetron sputtering, enabling scalability to large areas [12]. Owing to this unique combination of properties, YHO has emerged as a promising candidate for smart window applications.

Despite their promise, YHO films still face several challenges that must be addressed for practical applications. Environmental stability is one concern, as prolonged exposure to ambient conditions [13] and illumination [14] can drive hydrogen loss from the compound, leading to gradual deterioration of the photochromic response and, in some cases, mechanical failure such as delamination from the substrate [15]. Photochromic films also exhibit a characteristic yellowish appearance, which may limit their applicability in certain glazing contexts. In addition, slow bleaching kinetics are often observed, with recovery times extending from hours to days [16].

YHO thin films are predominantly synthesised by reactive magnetron sputtering using a two-step approach [3–5], in some cases complemented by post-deposition annealing [6], encapsulation [14,17], and anion labelling [7,18]. In the first step, a metallic yttrium target is sputtered in a hydrogen-containing atmosphere to form $\beta$-YH$_2$. In the second step, this hydride is oxidised, yielding a transparent and photochromic YHO phase [4]. Because the film microstructure governs both the oxidation process and the resulting optical – particularly

photochromic – properties, systematic studies across a wide range of deposition conditions are essential to optimise performance and to clarify the role of microstructure in the photochromic mechanism.

High power impulse magnetron sputtering (HiPIMS) is a sputtering technique in which a significant fraction of the sputtered flux is ionised [19]. It is a pulsed process that employs pulses (5–500 µs) at frequencies below 5 kHz, with duty cycles below 5 % and current densities reaching several A cm$^{-2}$. These intense pulses generate a dense plasma with particle densities of $10^{18}$–$10^{19}$ m$^{-3}$, enabling electron-impact ionisation of sputtered target atoms, in contrast to the predominantly Penning ionisation of Ar species occurring in conventional direct current magnetron sputtering (DCMS) discharges.

By comparison, pulsed-DCMS operates in the mid-frequency range of 20–350 kHz with duty cycles typically between 50 and 90 % [20]. During a defined reverse-time interval within each pulse, a positive reverse voltage – usually corresponding to approximately 10–15 % of the operating voltage, depending on the power supply – is applied. In pulsed-DCMS, peak current densities are at least two orders of magnitude lower than those achieved in HiPIMS, resulting in plasma densities not exceeding $10^{16}$ m$^{-3}$ [21].

Ion bombardment during growth enhances adatom mobility and improves adhesion by filling active bonding sites, typically resulting in dense, smooth, and homogeneous coatings with reduced porosity. HiPIMS has already demonstrated significant advantages in thin-film engineering, including precise control of microstructure [22], enhanced mechanical properties [23], improved performance at low deposition temperatures [24], and uniform coverage even on complex or non-flat substrates [25]. While HiPIMS is now well established in the hard coatings industry, its unique capabilities make it a promising method for the deposition of optoelectronic [26] and catalytic [27] thin films.

Applying HiPIMS to YHO deposition is expected to yield films with microstructural characteristics distinct from those produced by conventional sputtering, thereby influencing oxidation behaviour and, ultimately, the photochromic response. In the present study, YHO films were deposited using both HiPIMS and pulsed-DCMS over a range of sputtering pressures, and their optical – particularly photochromic – properties were analysed in relation to film growth conditions, microstructure, and composition. By employing different growth regimes, our study partially decouples the influence of film density on oxidation and provides insight into the way by which the deposition method governs the photochromic behaviour of YHO thin films.

**Experimental details**

YHO films were synthesized in a two-step process using a custom-built vacuum PVD coater (G500M, Sidrabe Vacuum, Ltd). In the first step, $\beta$-$YH_2$ films were deposited onto soda-lime glass and silicon (Si) substrates by reactive pulsed-DCMS (80 kHz, 2.5 µs reverse time with a positive voltage of 10 % of the operating voltage; power supply: EnerPulse 5, EN Technologies) and unipolar HiPIMS (140 Hz, 50 µs pulse length; power supply: SIP2000USB-10-500-D, Melec). In both cases, a Y target (99.9 % purity) was sputtered reactively in an Ar (99.9999 %) and $H_2$ (99.999 %) atmosphere, maintaining an $H_2$:Ar flow ratio of 2:15. In the second step, the chamber was vented to atmospheric pressure with air, which oxidised the films to form the YHO phase.

A planar balanced magnetron (150 mm × 75 mm × 3 mm) was mounted 11 cm from the grounded substrate holder, with substrates facing the target axis. No intentional substrate heating was applied during the deposition process. Prior to sputtering, the chamber was evacuated to $6 \times 10^{-4}$ Pa using a turbomolecular pump backed by a rotary pump. The target was operated at a constant time-averaged power of 200 W in both pulsed-DCMS and HiPIMS mode. Depositions were carried out at working pressures ranging from 0.45 to 1.20 Pa by adjusting the pumping speed with a throttle valve. Because the sputtering regime and pressure affect the deposition rate, deposition times were adjusted between 11 and 50 min to obtain films with thicknesses of approximately 500–600 nm. The deposition parameters for all samples are summarised in Table 1.

Table 1. Deposition parameters of YHO films prepared by reactive HiPIMS and pulsed-DCMS. The table lists the deposition regime, sample labels, sputtering pressure, discharge voltage, and resulting film thickness.

| Regime | Sample | Pressure (Pa) | Voltage (V) | Deposition time (min) | Thickness (nm) |
|---|---|---|---|---|---|
| HiPIMS | HiP1.00 | 1.00 | 533 | 46 | – |
|  | HiP1.05 | 1.05 | 522 | 50 | 577 |
|  | HiP1.10 | 1.10 | 514 | 50 | 550 |
|  | HiP1.15 | 1.15 | 517 | 50 | 552 |
|  | HiP1.20 | 1.20 | 513 | 50 | 557 |
| pulsed-DCMS | pDC0.45 | 0.45 | 280 | 15 | – |
|  | pDC0.55 | 0.55 | 273 | 15 | 606 |
|  | pDC0.60 | 0.60 | 273 | 12 | 505 |
|  | pDC0.70 | 0.70 | 272 | 12 | 542 |
|  | pDC0.80 | 0.80 | 265 | 11 | 531 |

The HiPIMS discharge voltage–current waveforms were measured using an MS-500-D-TB monitoring system (Melec) and recorded with a RIGOL DS1074B digital oscilloscope. Plasma optical emission spectra (OES) were collected through an optical fibre probe positioned inside the chamber, parallel to and approximately 1 cm above the target surface. Time-averaged OES from both pulsed-DCMS and HiPIMS discharges in the wavelength range of 300–1000 nm were measured using a PLASUS EMICON MC spectrometer. Time-resolved OES of individual emission lines were recorded during HiPIMS operation with an acousto-optic AOS 4-1 UV–Vis spectrometer (temporal resolution 0.1 µs; IFU GmbH) synchronised with the HiPIMS power supply. To improve signal quality, each emission line was measured three to five times during the HiPIMS pulse, and the resulting data were averaged.

As-deposited $β$-YH$_2$ films oxidise rapidly upon exposure to air, leading to a dramatic increase in visible-light transmittance and the formation of the YHO phase until a metastable state is reached. However, gradual continued oxidation and hydrogen release have been reported to occur in YHO films for up to seven months after deposition [13]. Thin-film characterisation – including X-ray diffraction (XRD), scanning electron microscopy (SEM), time-of-flight energy elastic recoil detection analysis (ToF-E ERDA), optical spectroscopy, spectroscopic ellipsometry (SE), and photochromic measurements – was carried out on samples in the metastable state. This state was typically reached within one week of air exposure and was identified by repeated transmittance measurements, which became constant once the metastable condition was established (see Fig. S1 in Supplementary Information - SI).

The light transmittance of the thin films in the wavelength range of 250–2500 nm was measured using an Agilent Cary 7000 spectrophotometer. Film thicknesses and refractive indexes ($n$) were determined using a WOOLLAM RC2 spectroscopic ellipsometer in the spectral range of 0.73–2.25 eV, where the films exhibit high transparency. The Cauchy model, $n(\lambda) = A + B\lambda^{-2}$, was applied in this energy range, with both the film thickness and optical constants extracted by fitting the measured $\Psi$ and $\Delta$ spectra using model-based regression implemented in the WOOLLAM CompleteEASE software [28]. The mean squared error (MSE) values for the models ranged between 7 and 15.

The composition analysis of selected films was performed using time-of-flight energy elastic recoil detection analysis (ToF-E ERDA). Measurements were carried out at the 5 MV tandem accelerator at the Helsinki Accelerator Laboratory (University of Helsinki) using a 50 MeV $^{127}I^{9+}$ ion beam as the probe. The samples were mounted with the surface normal rotated by 20° with respect to the incident beam, while the ToF telescope and energy detectors were positioned at 40° with respect to the direct beam. The data were recorded from the two detectors in the coincidence mode [29]. The resulted spectra were analysed by Potku software [30].

The crystal structure of the deposited films was analysed by XRD using a Rigaku MiniFlex 600 powder diffractometer in Bragg-Brentano $\theta$–$2\theta$ geometry with a 600 W Cu anode (Cu K$\alpha$ radiation, $\lambda$ = 1.54 Å). Crystallite size was estimated using the Scherrer equation, $\tau = K\lambda\,(\beta \cos \theta)^{-1}$, where $\tau$ is the crystallite size, $K$ is the shape factor (taken as 0.9), $\lambda$ is the X-ray wavelength, $\beta$ is the full width at half maximum of the diffraction peak in radians, and $\theta$ is the Bragg angle. To obtain the intrinsic peak broadening, $\beta$ was corrected for instrumental broadening (0.04°) according to $\beta = \beta_{exp} - \beta_{inst}$. Crystallite sizes were determined from the (111) and (200) reflections and averaged. Microstrain contributions were not subtracted, as Williamson–Hall analysis did not yield a clear linear dependence. Consequently, the reported crystallite sizes should be regarded as first-order estimates [31].

The surface and cross-sections of the films were examined using a Helios 5 UX dual-beam scanning electron microscope (SEM, Thermo Fisher Scientific). Cross-sectional images were obtained by scratching the films from the Si substrate and imaging the detached fragments positioned approximately perpendicular to the substrate surface.

The photochromic properties of the thin films were measured using a custom-built device comprising collecting and collimating probes, a spectrometer (Avantes StarLine AvaSpec-ULS2048CL-EVO), trigger and probing light sources, a timer to control the trigger lamp, and a sample holder (see Fig. S2 in SI). Photochromic darkening was induced under ambient conditions using a 15 W UVA–violet lamp ($\approx$2.4 mW cm$^{-2}$). The emission had a maximum at energy of 3.3 eV with a full width at half maximum (FWHM) of 0.13 eV. A tungsten-halogen lamp (Avantes AvaLight-HAL-S MINI) with a power density of $\approx$3.3 mW cm$^{-2}$ was used as the probing light.

## 3. Results and discussion

### 3.1. Yttrium sputtering in HiPIMS

Before depositing $\beta$-YH$_2$ thin films, the discharge properties of pulsed-DCMS and HiPIMS during sputtering of target in and Ar-H$_2$ atmosphere were characterised. A comparison of the time-averaged plasma OES for pulsed-DCMS (low plasma density) and HiPIMS (high plasma density) is shown in Fig. 1a. In the pulsed-DCMS discharge, the OES is dominated by neutral excited argon (Ar$^*$) and yttrium (Y$^*$) lines, with weak contributions from ionised argon (Ar$^+$). In contrast, under HiPIMS conditions the intensities of Ar$^*$ and Y$^*$ lines decrease markedly, while highly intense Y$^+$ lines dominate, reflecting the high degree of ionisation of sputtered Y atoms. The reduction of Ar-related emission is consistent with argon rarefaction – a localised depletion of working gas in front of the target caused by discharge heating, the hot target surface, momentum transfer from sputtered atoms (and reflected Ar atoms), and electron-impact ionisation [32,33].

The voltage–current waveforms and time-resolved emission of excited Ar$^*$ (751.5 nm), Y$^*$ (410.2 nm), H$^*$ (656.3 nm), together with ionised Ar$^+$ (434.8 nm) and Y$^+$ (437.5 nm), during a 50 μs HiPIMS pulse at 0.6 Pa are shown in Fig. 1b. For the other pressures used in thin-film deposition, similar voltage–current waveforms were observed. The OES line intensities were normalised because their absolute values span a wide range, with Y$^+$ being significantly more intense than the other lines. Due to the extremely low intensities of Ar$^*$ and H$^*$ lines during HiPIMS, the signals are noisy even after averaging.

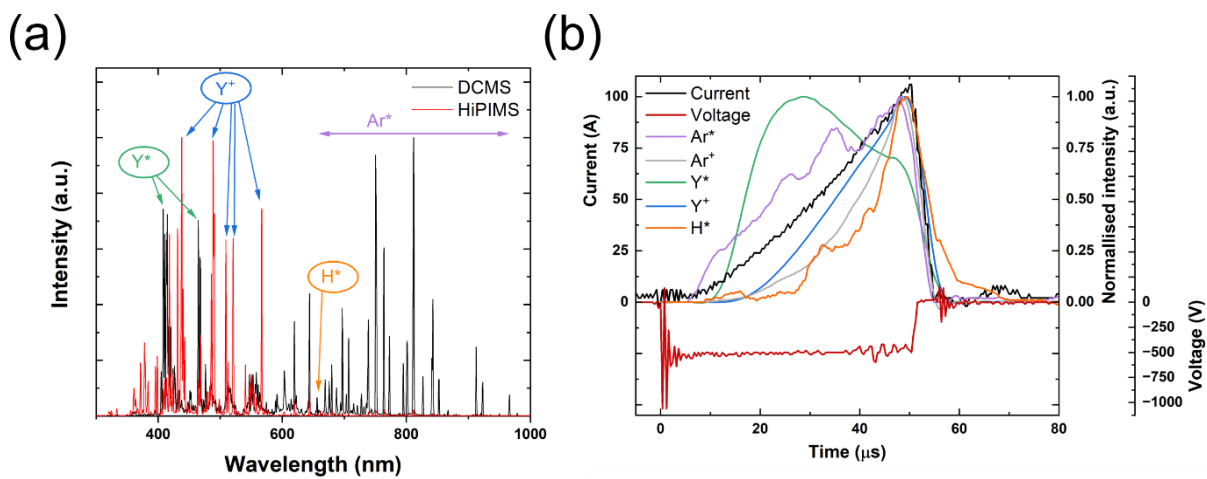

**Fig. 1.** (a) Time-averaged plasma emission spectra of pulsed-DCMS (80 kHz) and HiPIMS (140 Hz) discharges at 0.6 Pa, H$_2$:Ar flow ratio 2:15, and an average power density of 1.8 W

cm$^{-2}$. (b) Voltage–current waveforms and time-resolved emission intensities of Ar$^*$ (751.5 nm), Ar$^+$ (434.8 nm), Y$^*$ (410.2 nm), Y$^+$ (437.5 nm), and H$^*$ (656.3 nm) recorded during a 50 µs HiPIMS pulse.

At the beginning of the pulse, there is a ~5 µs delay between the applied voltage and the current rise, during which the plasma is negligible. Once ignition occurs, the voltage stabilises at ~515 V for the remainder of the pulse. The subsequent rapid increase in current arises because electron production exceeds recombination losses [34]. Energetic electrons trapped by the magnetic field near the target enhance ionisation of the working gas, reducing the plasma impedance as the charge-carrier density increases. This initial current rise is dominated by gas ions [35], with electrons gaining energy via sheath acceleration and Ohmic heating [36,37].

The onset of Ar$^+$, Y$^*$, and H$^*$ emission at ~9 µs indicates the start of sputtering. Y atoms sputtered from the target are subsequently ionised in the dense plasma, with Y$^+$ emission appearing a few microseconds later (~12 µs). A fraction of these Y$^+$ ions is drawn back to the target by the negative potential [38], which reduces the deposition rate to typically 30–85% of that in DCMS [39]. In our case, achieving the same film thickness required extending the HiPIMS deposition time by a factor of 3.3–4.5.

The current increases throughout the 50 µs pulse, forming a triangular profile with a peak of 110 A (0.98 A cm$^{-2}$) at the end, characteristic of HiPIMS regime [40]. Emission intensities of Ar$^*$, Ar$^+$, Y$^+$, and H$^*$ follow this current evolution and decay ~7 µs after the pulse ends. The triangular current rise without in-pulse decay is attributed to the relatively short pulse duration; in longer pulses, strong argon rarefaction typically leads to a mid-pulse reduction of Ar emission lines and a corresponding current decay [41].

The behaviour of the Y$^*$ line differs: its intensity peaks at ~28 µs and then declines, consistent with the progressive ionisation of Y atoms to Y$^+$. For other metals like Al and Ti, the ionised flux fraction (the ratio of ion flux to total particle flux at the substrate or detector [38]) reaches ≥50 % under typical HiPIMS conditions (current density ~1 A cm$^{-2}$, pressure 0.5–2 Pa) [42]. Given ionisation energy of Y (6.2 eV) is comparable to Al (6.0 eV) and Ti (6.8 eV), but with a larger electron-impact ionisation cross-section [43], the ionisation fraction for Y is expected to be similar or higher. Studies of ion energy distributions [44] have shown that Y$^+$ exhibits two components: a low-energy peak (~2 eV, thermalised ions) and a higher-energy peak (~8–10 eV, partially thermalised sputtered ions), with a high-energy tail extending up to ~100 eV.

By contrast, Ar$^+$ shows only a ~2 eV low-energy peak with a tail up to ~20 eV in the unipolar HiPIMS regime (1.25 A cm$^{-2}$, 1.85 Pa, 100 µs pulses at 100 Hz). The high-energy Y$^+$ component develops early in the pulse, while the thermalised component appears later.

The critical current for our sputtering process (0.6 Pa, target area 110 cm$^2$) is ~14 A, corresponding to a critical current density of 0.12 A cm$^{-2}$, as defined in Ref. [45]. This represents the maximum ion current achievable if all incoming Ar gas atoms are singly ionised and accelerated towards the target, considering the thermal refill rate of the working gas. Higher current densities require recycling mechanisms, either through ionised gas returning to the target or self-sputter recycling, or a combination of both. The balance between these processes depends on the self-sputter yield ($Y_{ss}$) [46]. For example, Al ($Y_{ss} \approx 1.0$ at 500 V) is strongly dominated by self-sputter recycling, whereas Ti ($Y_{ss} \approx 0.6$–0.7 at 600 V) lies at the boundary between self-sputter and mixed regimes [36]. To the best of our knowledge, $Y_{ss}$ has not been reported for yttrium. However, since sputter yields are primarily determined by the target material – or more precisely by the surface binding energy – $Y_{Ar}$ and $Y_{ss}$ are expected to be similar, differing by only ~10–15 % [47]. $Y_{Ar}$ for Y, approximately 0.75 at 515 V (see Fig. S3 in SI), is slightly higher than that for Ti. This indicates that self-sputter recycling contributes significantly to the discharge current in the case of Y. It is worth noting that the actual sputter yield values reported here for Y, Al, and Ti at the given discharge voltages are somewhat lower in practice, as Ar ions typically gain about 75–80 % of the total potential drop – some ions are generated within the sheath and therefore do not traverse the full potential [48].

### 3.2. Optical properties of YHO films

The sputtering pressure during $\beta$-YH$_2$ deposition is a key parameter, as previously shown in Ref. [49,50] for common pulsed-DCMS. Films deposited below the critical pressure ($P_c$) develop a microstructure that is too dense to permit sufficient oxygen incorporation during post-oxidation, preventing full conversion to the transparent YHO phase. Conversely, films deposited at pressures significantly above $P_c$ exhibit higher oxygen-to-yttrium ratios and greater transmittance, but at the expense of photochromic performance. Consequently, the highest photochromic contrast is achieved in films deposited slightly above $P_c$ [50].

To determine the $P_c$ values, the sputtering pressure was systematically varied between 0.45 and 0.80 Pa for pulsed-DCMS and between 1.00 and 1.20 Pa for HiPIMS (see Table 1). From Fig. 2a, the $P_c$ values were determined to be approximately ≳ 0.5 Pa and ≳ 1.0 Pa for pulsed-DCMS and HiPIMS, respectively, where the luminous and solar transmittance ($T_{\text{lum}}$ and $T_{\text{sol}}$) of films grown on soda-lime glass are plotted as a function of deposition pressure. These results indicate that higher sputtering pressures are required in the HiPIMS regime than in pulsed-DCMS to ensure sufficient oxygen uptake and the formation of transparent films. This behaviour arises because HiPIMS films are generally denser than those produced by pulsed-DCMS at equivalent deposition pressures [39]. It should also be noted that $P_c$ values are expected to depend on the specific deposition configuration, including the system geometry, the type of magnetron, and the target–substrate distance. For films deposited below $P_c$ (pDC0.45 and HiP1.00), the $T_{\text{sol}}$ values were significantly lower, reaching only 0.4 % and 4.0 %, respectively.

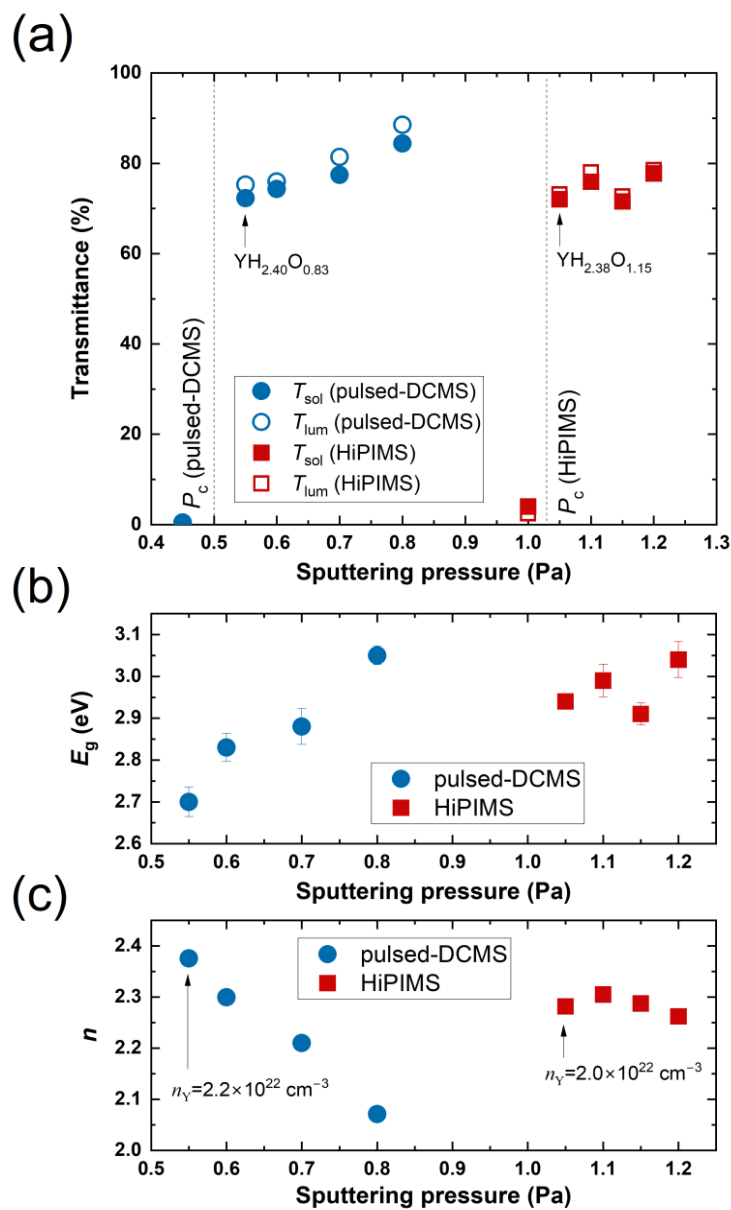

**Fig. 2.** (a) Solar ($T_{sol}$) and luminous ($T_{lum}$) transmittance, (b) optical band gap ($E_g$), and (c) refractive index ($n$) of 500–600 nm thick YHO films deposited on glass by reactive pulsed-DCMS and HiPIMS at varying sputtering pressures. Vertical dashed lines indicate the critical pressure ($P_c$) for each technique. In panels (a) and (c), the compositions and corresponding yttrium atomic densities ($n_Y$) of representative samples deposited slightly above $P_c$ (pDC0.55 and HiP1.05), determined by ToF-E ERDA, are indicated.

Above $P_c$, the transmittance continues to increase with pressure, from 72 % to 84 % ($T_{sol}$, 0.5–1.0 Pa) for pulsed-DCMS and from 72 % to 78 % ($T_{sol}$, 1.0–1.2 Pa) for HiPIMS. The solar and luminous transmittances are fundamental parameters for evaluating smart window performance and were calculated using the following equation:

$$T_{sol} = \frac{\int T(\lambda)\varphi_{sol}(\lambda)d\lambda}{\int \varphi_{sol}(\lambda)d\lambda},$$

where $\varphi_{sol}(\lambda)$ represents the solar irradiance spectrum at the Earth's surface. For luminous transmittance, $T_{sol}$ and $\varphi_{sol}(\lambda)$ are replaced by $T_{lum}$ and $\varphi_{lum}(\lambda)$, which represents the relative spectral sensitivity of the human eye [51].

Increasing the deposition pressure above $P_c$ results in a systematic increase in the optical band gap ($E_g$) of the YHO films, ranging from 2.70 to 3.05 eV for pulsed-DCMS and from 2.94 to 3.04 eV for HiPIMS (Fig. 2b). The $E_g$ values were determined from Tauc plots of $(\alpha h\nu)^2$ versus $h\nu$, assuming direct optical transitions, with $\alpha$ calculated from transmittance data (see Fig. S4). Despite exhibiting nearly identical optical transmittance, the transparent films deposited slightly above $P_c$ – HiP1.05 and pDC0.55 – display markedly different $E_g$ values of 2.94 eV and 2.70 eV, respectively.

The optical band gap of YHO is closely related to its composition, with a general tendency to increase with oxygen content [52]. This trend is consistent with the compositions of samples HiP1.05 and pDC0.55 determined by ToF-E ERDA, as shown in Fig. 2a. Although both films were deposited under optimal conditions (slightly above $P_c$), the pDC0.55 film exhibits a lower oxygen-to-hydrogen atomic ratio ($YH_{2.40}O_{0.83}$) compared with the HiP1.05 film ($YH_{2.38}O_{1.15}$). The corresponding compositional depth profiles are presented in Fig. S5. Both oxygen and hydrogen concentrations show gradients near the film surface, which are likely influenced by the film microstructure and exposure to ambient conditions; therefore, the reported compositions were extracted from deeper regions of the films where the elemental concentrations are uniform.

The ERDA measurements provide the areal atomic density (at cm$^{-2}$), which, in combination with the independently measured film thickness, was used to determine the volumetric atomic density (at cm$^{-3}$). The corresponding mass density was then calculated from the atomic density and employed as an input parameter in the Potku software to obtain depth profiles on a nanometre scale. This analysis reveals that the HiPIMS films possess a lower density of approximately 3.65 g cm$^{-3}$ compared with the pulsed-DCMS films, which exhibit a density of approximately 3.90 g cm$^{-3}$. The corresponding total atomic densities ($n_{at}$) and yttrium atomic densities ($n_Y$) are 9.1 × 10$^{22}$ cm$^{-3}$ (2.0 × 10$^{22}$ cm$^{-3}$) for the HiPIMS film and 9.5 × 10$^{22}$ cm$^{-3}$ (2.2 × 10$^{22}$ cm$^{-3}$) for the pulsed-DCMS film, respectively (see Fig. 2c). This difference is attributed to the higher sputtering pressures employed during HiPIMS deposition; although HiPIMS generally produces denser films than pulsed-DCMS, this trend does not necessarily hold when the two techniques are compared at different deposition pressures. The lower density of the HiPIMS film also provides a explanation for its higher oxygen-to-hydrogen atomic ratio.

The refractive index ($n$) primarily depends on both the composition and the density of the films. SE modelling revealed an $n$ gradient through the film thickness, decreasing from the substrate towards the surface, consistent with previous observations [4]. The average $n$ values across the film thickness at 2.25 eV are plotted in Fig. 2c. Because composition and density in YHO films are intrinsically coupled by the deposition process, it is difficult to identify which of these factors predominantly governs the higher $n$ observed in pDC0.55 relative to HiP1.05 deposited slightly above $P_c$.

For pulsed-DCMS films, $n$ decreases markedly from 2.37 to 2.07 with increasing sputtering pressure, whereas in HiPIMS films the reduction in $n$ with pressure is much less pronounced, or even negligible. This clearly indicates that pulsed-DCMS films are more sensitive to variations in sputtering pressure. Overall, considering optical properties alone, the HiPIMS films resemble pulsed-DCMS films deposited at pressures of approximately 0.6–0.7 Pa.

### 3.3. Structure of YHO films

YHO films crystallise in a face-centred cubic (*fcc*) cation lattice [53], where oxygen anions occupy the vacant tetrahedral sites, while hydrogen can occupy both tetrahedral and octahedral positions according to theoretical predictions [54]. Computational studies further suggest that a random anionic distribution within the tetrahedral sites is energetically more favourable than an ordered configuration [55], a conclusion supported by Fourier-transform infrared (FTIR) and solid-state nuclear magnetic resonance (NMR) analyses [18]. Moreover, structural relaxation may allow hydrogen anions to occupy intermediate positions between ideal tetrahedral and octahedral sites.

XRD patterns of representative samples deposited slightly above $P_c$ – HiP1.05 and pDC0.55 – are shown in Fig. 3a, while the complete set of diffractograms is provided in Fig. S6 (SI). Both films exhibit the characteristic diffraction peaks of the *fcc* lattice: (111), (200), (220), and (311). These peaks are shifted to lower diffraction angles compared to *β*-YH$_2$ (ICDD 04-002-6939), indicating lattice expansion due to oxygen incorporation [56].

The most pronounced structural difference between the two deposition regimes lies in the relative intensities of the (111) and (200) reflections. The pulsed-DCMS films exhibit a strong out-of-plane preferential orientation along the <100> direction, whereas the HiPIMS films display a more weakly textured, nearly random polycrystalline orientation, with relative peak intensities approaching those expected for a powder diffraction pattern (see Fig. 3a for reference peak intensities). Given that the (111) surface is expected to have the lowest surface energy in YH$_2$ of 0.78 J m$^{-2}$ [57], the enhanced (111) contribution in the HiPIMS films is consistent with a growth mode closer to thermodynamic equilibrium during energetic HiPIMS deposition. In contrast, the pronounced <100> texture observed in the pulsed-DCMS films suggests growth under more kinetically limited conditions, where anisotropies in surface diffusion and limited adatom mobility may favour the non-equilibrium orientation. Quantitative data on diffusion barriers for YH$_2$ are currently scarce; however, analogous behaviour has been reported for *fcc* structures [58], in which growth orientation is governed by the balance between surface-energy minimisation and anisotropic surface diffusion. The only exception is the opaque pDC0.45 film deposited below $P_c$, which displays an XRD pattern similar to that of the HiPIMS films (see Fig. S6). This shows that, for pulsed-DCMS films, $P_c$ represents a threshold above which the film microstructure changes significantly, strongly influencing the subsequent oxidation through oxygen diffusion along grain boundaries and intercolumnar voids.

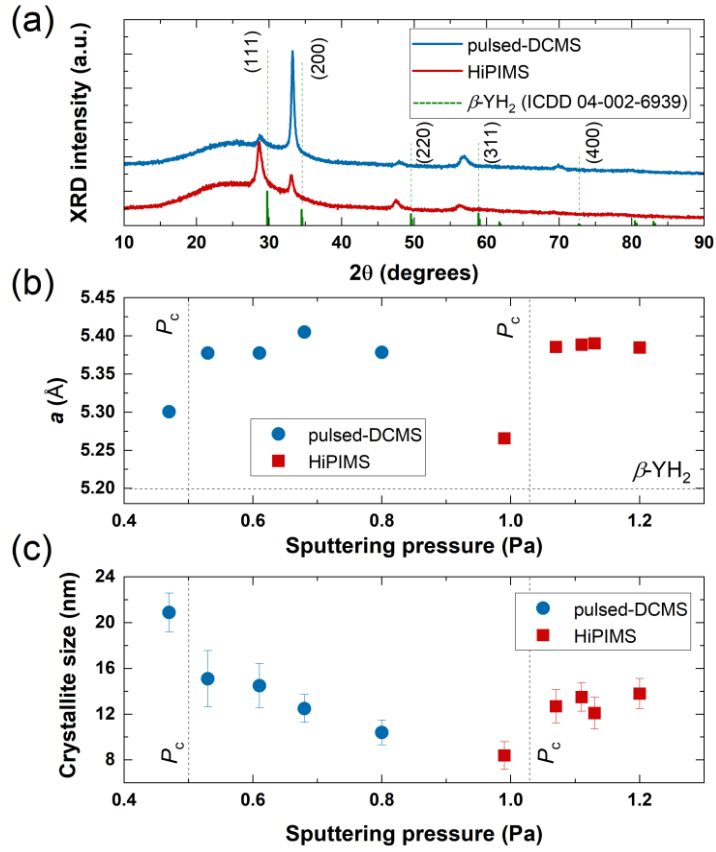

**Fig. 3.** (a) XRD patterns of YHO films deposited slightly above $P_c$ by pulsed-DCMS (sample pDC0.55) and HiPIMS (sample HiP1.05). Reference peak positions for $\beta$-YH$_2$ (ICDD 04-002-6939) are indicated by vertical markers. (b) Lattice parameter $a$ as a function of sputtering pressure for both deposition regimes; the horizontal dashed line indicates the value of $a$ (5.2 Å) for $\beta$-YH$_2$. (c) Crystallite size ($\tau$) as a function of sputtering pressure. Vertical dashed lines denote $P_c$ values for each technique.

To enable a more consistent comparison of lattice parameters and crystallite sizes across samples, the average values derived from the (111) and (200) peaks were used. For both deposition regimes, a sudden increase in lattice parameter $a$ is observed once the sputtering pressure exceeds $P_c$ – from 5.30 Å (pulsed-DCMS) and 5.27 Å (HiPIMS) to 5.38–5.41 Å (Fig. 3b). Above $P_c$, no clear correlation was observed between $a$ and the pressure. In contrast, the evolution of crystallite size ($\tau$) with pressure exhibits an opposite trend for the two regimes (Fig. 3c): $\tau$ decreases from 21 nm to 10 nm for pulsed-DCMS but increases from 8 nm to 14 nm for HiPIMS. The smaller crystallite sizes near $P_c$ in the HiPIMS films may additionally contribute to their slightly larger $E_g$ values compared with the pulsed-DCMS films, owing to quantum confinement effects discussed in Ref. [10].

The general trend of decreasing crystallite size with increasing total sputtering pressure is well established and attributed to reduced adatom mobility and enhanced gas-phase scattering of sputtered species, consistent with the behaviour observed in pulsed-DCMS. However, in energetic deposition techniques such as HiPIMS, where the ion flux and bombardment energy strongly influence adatom diffusion and film microstructure, this pressure-induced grain refinement can be counteracted. In our experiments, a slight increase in the average discharge current (from 0.36 A to 0.38 A) was accompanied by a corresponding drop in voltage (from 533 V to 513 V), since the deposition was performed at constant average power, as the pressure was raised from 1.0 to 1.2 Pa. Because the degree of ionisation and the kinetic energy of ions are closely linked to the peak current and discharge voltage [59], these results suggest that the conditions at higher pressures favour the growth of slightly larger crystallites.

Top-view and cross-sectional SEM images of YHO films deposited by pulsed-DCMS (pDC0.60) and HiPIMS (HiP1.05) are shown in Fig. 4. Both films exhibit densely packed, fine-grained surface morphologies, although the surface features of the HiPIMS film appear slightly smaller. Cross-sectional images reveal columnar, fibre-like grains that widen with film growth. The overall morphology of the pulsed-DCMS and HiPIMS films appears broadly similar, suggesting that both deposition regimes yield comparable columnar structures under the present conditions. Any subtle differences in surface texture or column uniformity may arise from the distinct energy fluxes and ion bombardment characteristics of the two techniques.

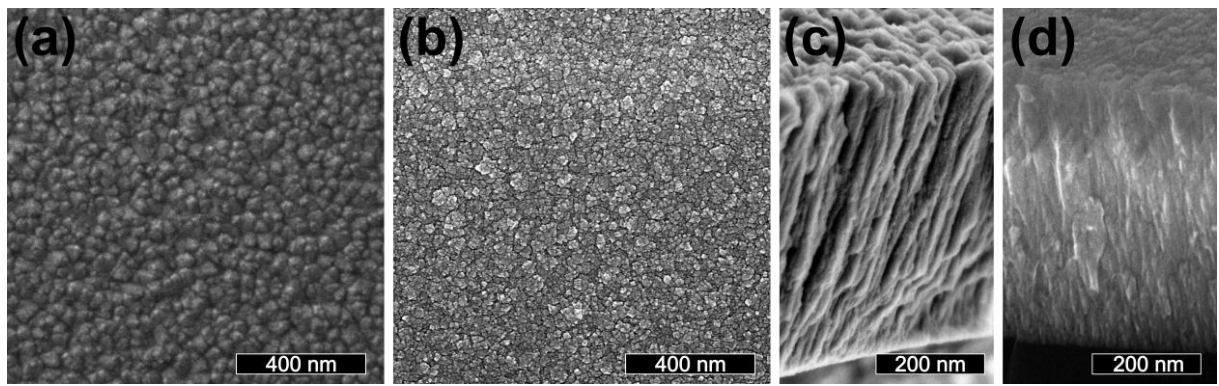

**Fig. 4.** Surface and cross-sectional SEM images of YHO films: (a,c) pulsed-DCMS sample pDC0.60 and (b,d) HiPIMS sample HiP1.05.

## 3.4. Photochromic properties of YHO

The photochromic response of the films was evaluated by illuminating the films with a UVA–violet lamp for one hour (see Fig. S7). The relative contrast ($\Delta T_r = \Delta T / T_0$, where $T_0$ is the transmittance before illumination), averaged over the 500–700 nm wavelength range, is plotted in Fig. 5a. For both HiPIMS and pulsed-DCMS samples, increasing the sputtering pressure above $P_c$ results in a gradual reduction in photochromic contrast, consistent with an increased oxygen content that exceeds the optimum level for photochromic activity [5,6]. Films deposited by pulsed-DCMS exhibit a significantly higher contrast ($\Delta T_r$ = 19–34 %) compared with HiPIMS films ($\Delta T_r$ = 5–9 %). Based on the compositions measured by ToF-E ERDA for representative samples deposited slightly above $P_c$, the pulsed-DCMS film contains slightly less oxygen than its HiPIMS counterpart (see Fig. 2a), which partly explains its significantly higher $\Delta T_r$. Notably, even at the highest deposition pressure (0.80 Pa), the pulsed-DCMS sample (pDC0.8) exhibits a substantially higher photochromic contrast than the HiPIMS films, despite having a higher solar transmittance and optical band gap, a lower refractive index (Fig. 2), a similar lattice parameter, and a slightly smaller crystallite size (Fig. 3).

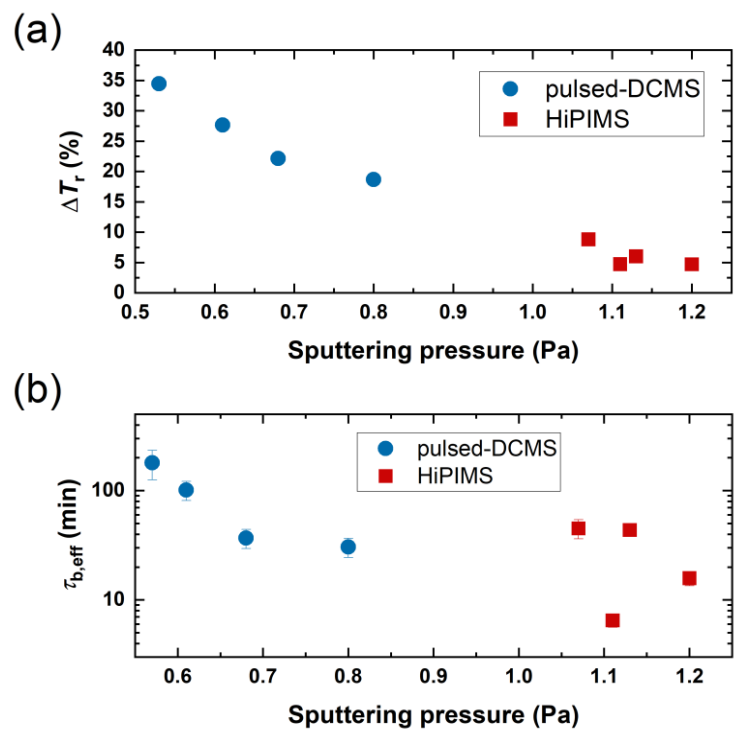

**Fig. 5.** (a) Relative photochromic contrast ($\Delta T_r$) and (b) effective bleaching time constant ($\tau_{b,eff}$) as a function of sputtering pressure for YHO films deposited by reactive pulsed-DCMS and HiPIMS.

From the perspective of the thin-film deposition process, HiPIMS films may undergo more pronounced partial oxidation during the YH$_2$ deposition step than films grown by pulsed-DCMS, which is highly undesirable, as shown in Ref. [4]. This effect can be attributed to two factors: (i) the higher deposition pressures in HiPIMS were achieved by reducing the pumping speed, thereby increasing the partial pressure of residual gases, and (ii) the deposition rate in HiPIMS was approximately 3.3–4.5 times lower, which prolongs the exposure time of the growing film to residual gases and increases the likelihood of unwanted reactions. This combination of increased residual gas exposure, longer growth times, and possible differences in target poisoning likely contributes to the observed differences in film composition between the two deposition regimes.

The origin of photochromism in YHO remains under debate, with several models proposed to explain the light-induced optical changes, including the formation of metallic domains [8,60] and anion-vacancy-mediated electronic transitions that modify local charge states and lattice structure [9].

Another factor distinguishing the two deposition regimes is the film texture. As density and composition in YHO films are intrinsically coupled, there is currently no detailed understanding of how density alone influences the photochromic properties. Likewise, no systematic studies have yet been conducted to assess how preferred crystallographic orientation affects photochromism in YHO. Among reports of photochromic YHO films ($\Delta T > 10$ %), the ratio of the (100) to (200) XRD peak intensities varies substantially between studies [61–63], and even within individual reports [4,64,65], with no clear correlation observed between texture and photochromic contrast. Further investigation is therefore required to clarify whether crystallographic orientation influences oxidation behaviour and the resulting magnitude of the photochromic effect.

The highest measured $\Delta T_r$ values for the pulsed-DCMS YHO films in this study are still lower than those reported in the literature, where contrasts of up to 45 % have been achieved after one hour of illumination [3,15]. This deviation is primarily attributed to the lower intensity of the illumination source used here, as the photochromic darkening is known to scale strongly with light power [66]. Additionally, factors such as film thickness and specific deposition conditions further influence the observed contrast, indicating scope for optimisation in future studies.

To quantify the rate of recovery (bleaching) of the films, the following bi-exponential relation was used [67]:

$$-\ln\left(\frac{T(t)}{T_0}\right) = A_1 e^{-t/\tau_{b1}} + A_2 e^{-t/\tau_{b2}}$$

where $T(t)$ is the transmittance as a function of time, $A_1$ and $A_2$ are the initial amplitudes associated with two distinct relaxation processes, and $\tau_{b1}$ (fast) and $\tau_{b2}$ (slow) are their respective bleaching time constants, respectively. The expression assumes that the decay of the photo-induced absorbing species follows first-order kinetics [68]. The bleaching time constants were extracted from fits to the logarithmic plots of $-\ln(T / T_0)$ versus time (see Fig. S8). To enable a direct comparison between samples (Fig. 5b), an effective bleaching time constant ($\tau_{b,\text{eff}}$) was calculated as an amplitude-weighted average of the two bleaching contributions,

$$\tau_{b,\text{eff}} = \frac{A_1 \tau_{b1} + A_2 \tau_{b2}}{A_1 + A_2}.$$

For HiPIMS films, $\tau_{b,\text{eff}}$ decreases from 45 min to 7 min with increasing sputtering pressure, while for pulsed-DCMS films it decreases from 180 min to 31 min, indicating that higher deposition pressures lead to faster bleaching. An inverse relationship between photochromic contrast and bleaching rate is evident: films exhibiting higher photochromic contrast recover more slowly. This trade-off highlights the need to balance photochromic contrast and bleaching kinetics when designing YHO-based smart window coatings. All the characteristics of the films investigated in this study are summarised in Table 2.

Table 2. Summary of the key optical, photochromic, and structural properties of YHO films deposited by HiPIMS and pulsed-DCMS, including thickness $d$, solar transmittance ($T_{sol}$), optical band gap ($E_g$), average refractive index ($n$) at 2.25 eV, relative photochromic contrast ($\Delta T_r$), effective bleaching time constant ($\tau_{b,eff}$), lattice parameter ($a$), and crystallite size ($\tau$). The compositions and both mass and atomic densities of representative samples deposited slightly above $P_c$ were determined by ToF-E ERDA and are indicated by asterisks.

| Sample | $d$ (nm) | $T_{sol}$ (%) | $E_g$ (eV) | $n$ | $\Delta T_r$ (%) | $\tau_{b,eff}$ (min) | $a$ (Å) | $\tau$ (nm) |
|---|---|---|---|---|---|---|---|---|
| HiP1.00 | – | 4.0 | – | – | – | – | 5.31 | 8 |
| HiP1.05* | 577 | 72.0 | 2.94 | 2.28 | 8.8 | 45.2 | 5.39 | 13 |
| HiP1.10 | 550 | 75.9 | 2.99 | 2.31 | 4.7 | 6.5 | 5.40 | 14 |
| HiP1.15 | 552 | 71.6 | 2.91 | 2.29 | 6.0 | 43.6 | 5.40 | 12 |
| HiP1.20 | 557 | 78.4 | 3.04 | 2.26 | 4.7 | 15.8 | 5.40 | 14 |
| pDC0.45 | – | 0.4 | – | – | – | – | 5.29 | 21 |
| pDC0.55** | 606 | 72.2 | 2.70 | 2.38 | 34.5 | 179.6 | 5.38 | 15 |
| pDC0.60 | 505 | 74.3 | 2.83 | 2.30 | 27.7 | 101.4 | 5.39 | 15 |
| pDC0.70 | 542 | 77.4 | 2.88 | 2.21 | 22.2 | 36.9 | 5.41 | 13 |
| pDC0.80 | 531 | 84.4 | 3.05 | 2.07 | 18.7 | 30.6 | 5.38 | 10 |

*$YH_{2.38}O_{1.15}$; $\rho = 3.65$ g cm$^{-3}$; $n_{at} = 9.1 \times 10^{22}$ cm$^{-3}$; $n_Y = 2.0 \times 10^{22}$ cm$^{-3}$
**$YH_{2.40}O_{0.83}$; $\rho = 3.90$ g cm$^{-3}$; $n_{at} = 9.5 \times 10^{22}$ cm$^{-3}$; $n_Y = 2.2 \times 10^{22}$ cm$^{-3}$

## Conclusions

Photochromic YHO thin films were deposited by reactive HiPIMS and pulsed-DCMS, and the influence of sputtering pressure on their optical, structural, and photochromic properties was systematically investigated. Optical emission spectroscopy shows that, in contrast to pulsed-DCMS discharges dominated by $Ar^+$ ions, HiPIMS discharges are characterised by strong $Y^+$ emission, indicating a high degree of yttrium ionisation and a substantial contribution from self-sputter recycling. To obtain transparent and photochromic films, a higher critical pressure was required for HiPIMS ($P_c \gtrsim 1.0$ Pa) than for pulsed-DCMS ($P_c \gtrsim 0.5$ Pa).

While both deposition techniques exhibit similar pressure-dependent trends in optical properties, substantial differences were observed in photochromic performance and texture. Pulsed-DCMS films show significantly higher relative photochromic contrast ($\Delta T_r$ = 19–34 %) than HiPIMS films ($\Delta T_r$ = 5–9 %). Compositional analysis of representative samples deposited near $P_c$ indicates a lower oxygen-to-hydrogen atomic ratio and higher Y atomic density in pulsed-DCMS films, which partly explains their enhanced photochromic response. The higher sputtering pressures and lower deposition rates during HiPIMS appear to constitute non-optimal growth conditions, resulting in reduced photochromic activity.

Structurally, pulsed-DCMS films exhibit a pronounced preferential <100> orientation, whereas HiPIMS films exhibit a random texture. Under the conditions explored here, HiPIMS is therefore not optimal for producing highly photochromic YHO films. However, further optimisation – through improved base vacuum conditions, higher deposition rates, and independent control of ion energy and ion flux towards the growing film – may enable enhanced photochromic performance in future HiPIMS-based deposition strategies.

## Declaration of competing interest

The authors declare that they have no known competing financial interests or personal relationships that could have appeared to influence the work reported in this paper.

## Acknowledgements

Financial support was provided by Latvian Council of Science Project No. lzp-2022/2-0454. ToF-E ERDA measurements by MZ were supported from post-doctoral research project 1.1.1.9/LZP/1/24/011. The research team acknowledges the SWEB project 101087367 funded by the HORIZON-WIDERA-2022-TALENTS-01, which provided support for competence development and training. The authors thank A. Ashraf for her assistance with the ToF-E ERDA measurements. KS acknowledges financial support by the Swedish Research Council (Grant No. VR-2021-04113) and the Åforsk foundation (Grant No. 22-150).

During the preparation of this work the author(s) used ChatGPT in order to enhance language clarity and grammatical accuracy. After using this tool/service, the author(s) reviewed and edited the content as needed and take(s) full responsibility for the content of the published article.

**Data availability**

Data will be made available on request.

# References

[1] Rezaei SD, Shannigrahi S, Ramakrishna S. A review of conventional, advanced, and smart glazing technologies and materials for improving indoor environment. Solar Energy Materials and Solar Cells 2017;159:26–51. https://doi.org/10.1016/j.solmat.2016.08.026.

[2] Ke Y, Chen J, Lin G, Wang S, Zhou Y, Yin J, et al. Smart Windows: Electro-, Thermo-, Mechano-, Photochromics, and Beyond. Adv Energy Mater 2019;9. https://doi.org/10.1002/aenm.201902066.

[3] Mongstad T, Platzer-Björkman C, Maehlen JP, Mooij LPA, Pivak Y, Dam B, et al. A new thin film photochromic material: Oxygen-containing yttrium hydride. Solar Energy Materials and Solar Cells 2011;95:3596–9. https://doi.org/10.1016/j.solmat.2011.08.018.

[4] Zubkins M, Aulika I, Strods E, Vibornijs V, Bikse L, Sarakovskis A, et al. Optical properties of oxygen-containing yttrium hydride thin films during and after the deposition. Vacuum 2022;203. https://doi.org/10.1016/j.vacuum.2022.111218.

[5] You CC, Moldarev D, Mongstad T, Primetzhofer D, Wolff M, Marstein ES, et al. Enhanced photochromic response in oxygen-containing yttrium hydride thin films transformed by an oxidation process. Solar Energy Materials and Solar Cells 2017;166:185–9. https://doi.org/10.1016/j.solmat.2017.03.023.

[6] Moldarev D, Moro M V, You CC, Baba EM, Karazhanov SZ, Wolff M, et al. Yttrium oxyhydrides for photochromic applications: Correlating composition and optical response. Phys Rev Mater 2018;00. https://doi.org/10.1103/PhysRevMaterials.00.005200.

[7] Banerjee S, Chaykina D, Stigter R, Colombi G, Eijt SWH, Dam B, et al. Exploring Multi-Anion Chemistry in Yttrium Oxyhydrides: Solid-State NMR Studies and DFT Calculations. Journal of Physical Chemistry C 2023;127:14303–16. https://doi.org/10.1021/acs.jpcc.3c02680.

[8] Dam B, Nafezarefi F, Chaykina D, Colombi G, Wu Z, Eijt SWH, et al. Perspective on the photochromic and photoconductive properties of Rare-Earth Oxyhydride thin films. Solar Energy Materials and Solar Cells 2024;273. https://doi.org/10.1016/j.solmat.2024.112921.

[9] Arslan H, Kuzmin A, Kumar Kasi V, Aulika I, Moldarev D, Primetzhofer D, et al. Anion vacancy-induced photochromism and lattice relaxation in yttrium oxyhydride. Commun Mater 2025;6. https://doi.org/10.1038/s43246-025-00868-2.

[10] Moldarev D, Wolff M, Baba EM, Moro M V., You CC, Primetzhofer D, et al. Photochromic properties of yttrium oxyhydride thin films: Surface versus bulk effect. Materialia (Oxf) 2020;11. https://doi.org/10.1016/j.mtla.2020.100706.

[11] You CC, Karazhanov SZ. Effect of temperature and illumination conditions on the photochromic performance of yttrium oxyhydride thin films. J Appl Phys 2020;128. https://doi.org/10.1063/5.0010132.

[12] Zubkins M, Fedotovs A, Vibornijs V, Rudevica Z, Spunde K, Strods E, et al. Scaling up photochromic thin films with antimicrobial functionality: Roll-to-Roll deposition of YHO and YHO/Cu coatings. Vacuum 2026;244:114892. https://doi.org/10.1016/j.vacuum.2025.114892.

[13]  Moldarev D, Primetzhofer D, You CC, Karazhanov SZ, Montero J, Martinsen F, et al. Composition of photochromic oxygen-containing yttrium hydride films. Solar Energy Materials and Solar Cells 2018;177:66–9. https://doi.org/10.1016/j.solmat.2017.05.052.

[14]  Strods E, Zubkins M, Vibornijs V, Moldarev D, Sarakovskis A, Kundzins K, et al. Role of hydrogen dynamics and deposition conditions in photochromic YHO/MoO3 bilayer films. Solar Energy Materials and Solar Cells 2025;292. https://doi.org/10.1016/j.solmat.2025.113789.

[15]  Montero J, Martinsen FA, Lelis M, Karazhanov SZ, Hauback BC, Marstein ES. Preparation of yttrium hydride-based photochromic films by reactive magnetron sputtering. Solar Energy Materials and Solar Cells 2018;177:106–9. https://doi.org/10.1016/j.solmat.2017.02.001.

[16]  Wu Z, De Wit L, Beek M, Colombi G, Chaykina D, Schreuders H, et al. Memory effect in photochromic rare-earth oxyhydride thin films studied by in situ positron annihilation spectroscopy upon photodarkening-bleaching cycling. Phys Rev Mater 2024;8. https://doi.org/10.1103/PhysRevMaterials.8.045201.

[17]  Moro M V., Aðalsteinsson SM, Tran TT, Moldarev D, Samanta A, Wolff M, et al. Photochromic Response of Encapsulated Oxygen-Containing Yttrium Hydride Thin Films. Physica Status Solidi - Rapid Research Letters 2021;15. https://doi.org/10.1002/pssr.202000608.

[18]  Zubkins M, Gabrusenoks J, Aleksis R, Chikvaidze G, Strods E, Vibornijs V, et al. Vibrational properties of photochromic yttrium oxyhydride and oxydeuteride thin films. J Alloys Compd 2025;1015. https://doi.org/10.1016/j.jallcom.2025.178917.

[19]  Sarakinos K, Alami J, Konstantinidis S. High power pulsed magnetron sputtering: A review on scientific and engineering state of the art. Surf Coat Technol 2010;204:1661–84. https://doi.org/10.1016/j.surfcoat.2009.11.013.

[20]  Bäcker H, Henderson PS, Bradley JW, Kelly PJ. Time-resolved investigation of plasma parameters during deposition of Ti and TiO2 thin films. Surf Coat Technol 2003;174–175:909–13. https://doi.org/10.1016/S0257-8972(03)00469-9.

[21]  Bradley JW, Bäcker H, Kelly PJ, Arnell RD. Time-resolved Langmuir probe measurements at the substrate position in a pulsed mid-frequency DC magnetron plasma. Surf Coat Technol 2001;135:221–8. https://doi.org/10.1016/S0257-8972(00)00990-7.

[22]  Konstantinidis S, Dauchot JP, Hecq M. Titanium oxide thin films deposited by high-power impulse magnetron sputtering. Thin Solid Films 2006;515:1182–6. https://doi.org/10.1016/j.tsf.2006.07.089.

[23]  Greczynski G, Lu J, Jensen J, Petrov I, Greene JE, Bolz S, et al. Strain-free, single-phase metastable Ti0.38Al0.62N alloys with high hardness: Metal-ion energy vs. momentum effects during film growth by hybrid high-power pulsed/dc magnetron cosputtering. Thin Solid Films 2014;556:87–98. https://doi.org/10.1016/j.tsf.2014.01.017.

[24]  Wallin E, Selinder TI, Elfwing M, Helmersson U. Synthesis of α-Al2 O3 thin films using reactive high-power impulse magnetron sputtering. EPL 2008;82. https://doi.org/10.1209/0295-5075/82/36002.

[25]  Alami J, Persson POÅ, Music D, Gudmundsson JT, Bohlmark J, Helmersson U. Ion-assisted physical vapor deposition for enhanced film properties on nonflat surfaces. Journal of

Vacuum Science & Technology A: Vacuum, Surfaces, and Films 2005;23:278–80. https://doi.org/10.1116/1.1861049.

[26] Zubkins M, Arslan H, Bikse L, Purans J. High power impulse magnetron sputtering of Zn/Al target in an Ar and Ar/O$_2$ atmosphere: The study of sputtering process and AZO films. Surf Coat Technol 2019;369. https://doi.org/10.1016/j.surfcoat.2019.04.044.

[27] Zubkins M, Sarakovskis A, Strods E, Bikse L, Polyakov B, Kuzmin A, et al. Tailoring of rhenium oxidation state in ReOx thin films during reactive HiPIMS deposition process and following annealing. Mater Chem Phys 2022;289. https://doi.org/10.1016/j.matchemphys.2022.126399.

[28] J.A. Woollam Co. Inc. CompleteEASE™ Spectroscopic Ellipsometry Software 2023.

[29] Whitlow HJ, Possnert G, Petersson CS. Quantitative mass and energy dispersive elastic recoil spectrometry: Resolution and efficiency considerations. Nucl Instrum Methods Phys Res B 1987;27:448–57. https://doi.org/10.1016/0168-583X(87)90527-1.

[30] Arstila K, Julin J, Laitinen MI, Aalto J, Konu T, Kärkkäinen S, et al. Potku – New analysis software for heavy ion elastic recoil detection analysis. Nucl Instrum Methods Phys Res B 2014;331:34–41. https://doi.org/10.1016/j.nimb.2014.02.016.

[31] Harrington GF, Santiso J. Back-to-Basics tutorial: X-ray diffraction of thin films. J Electroceram 2021;47:141–63. https://doi.org/10.1007/s10832-021-00263-6.

[32] Huo C, Raadu MA, Lundin D, Gudmundsson JT, Anders A, Brenning N. Gas rarefaction and the time evolution of long high-power impulse magnetron sputtering pulses. Plasma Sources Sci Technol 2012;21. https://doi.org/10.1088/0963-0252/21/4/045004.

[33] Rossnagel SM. Gas density reduction effects in magnetrons. Journal of Vacuum Science & Technology A: Vacuum, Surfaces, and Films 1988;6:19–24. https://doi.org/10.1116/1.574988.

[34] Wu Z, Xiao S, Ma Z, Cui S, Ji S, Tian X, et al. Discharge current modes of high power impulse magnetron sputtering. AIP Adv 2015;5. https://doi.org/10.1063/1.4932135.

[35] Anders A, Andersson J, Ehiasarian A. High power impulse magnetron sputtering: Current-voltage-time characteristics indicate the onset of sustained self-sputtering. J Appl Phys 2007;102. https://doi.org/10.1063/1.2817812.

[36] Huo C, Lundin D, Gudmundsson JT, Raadu MA, Bradley JW, Brenning N. Particle-balance models for pulsed sputtering magnetrons. J Phys D Appl Phys 2017;50. https://doi.org/10.1088/1361-6463/aa7d35.

[37] Brenning N, Gudmundsson JT, Lundin D, Minea T, Raadu MA, Helmersson U. The role of Ohmic heating in dc magnetron sputtering. Plasma Sources Sci Technol 2016;25. https://doi.org/10.1088/0963-0252/25/6/065024.

[38] Butler A, Brenning N, Raadu MA, Gudmundsson JT, Minea T, Lundin D. On three different ways to quantify the degree of ionization in sputtering magnetrons. Plasma Sources Sci Technol 2018;27. https://doi.org/10.1088/1361-6595/aae05b.

[39] Samuelsson M, Lundin D, Jensen J, Raadu MA, Gudmundsson JT, Helmersson U. On the film density using high power impulse magnetron sputtering. Surf Coat Technol 2010;205:591–6. https://doi.org/10.1016/j.surfcoat.2010.07.041.

[40] Gudmundsson JT, Brenning N, Lundin D, Helmersson U. High power impulse magnetron sputtering discharge. Journal of Vacuum Science & Technology A: Vacuum, Surfaces, and Films 2012;30. https://doi.org/10.1116/1.3691832.

[41] Ehiasarian AP, New R, Unz W-DM, Hultman L, Helmersson U, Kouznetsov V. Influence of high power densities on the composition of pulsed magnetron plasmas. vol. 65. 2002.

[42] Lundin Daniel, Minea Tiberiu, Gudmundsson JTomas. High power impulse magnetron sputtering : fundamentals, technologies, challenges and applications. Elsevier; 2020.

[43] Mann JB. Ionization cross sections of the elements calculated from mean-square radii of atomic orbitals. J Chem Phys 1967;46:1646–51. https://doi.org/10.1063/1.1840917.

[44] Hippler R, Cada M, Hubicka Z. Time-resolved diagnostics of a bipolar HiPIMS discharge. J Appl Phys 2020;127. https://doi.org/10.1063/5.0006425.

[45] Huo C, Lundin D, Raadu MA, Anders A, Gudmundsson JT, Brenning N. On the road to self-sputtering in high power impulse magnetron sputtering: Particle balance and discharge characteristics. Plasma Sources Sci Technol 2014;23. https://doi.org/10.1088/0963-0252/23/2/025017.

[46] Brenning N, Gudmundsson JT, Raadu MA, Petty TJ, Minea T, Lundin D. A unified treatment of self-sputtering, process gas recycling, and runaway for high power impulse sputtering magnetrons. Plasma Sources Sci Technol 2017;26. https://doi.org/10.1088/1361-6595/aa959b.

[47] Anders A. Deposition rates of high power impulse magnetron sputtering: Physics and economics. Journal of Vacuum Science & Technology A: Vacuum, Surfaces, and Films 2010;28:783–90. https://doi.org/10.1116/1.3299267.

[48] Czekaj D, Goranchev B, Hollmann E, Volpyas V, Zaytsev A. Incident ion energy spectrum and target sputtering rate in dc planar magnetron. Vacuum 1991;42:43–5. https://doi.org/10.1016/0042-207X(91)90075-T.

[49] Montero-Amenedo J. Photochromism in rare earth oxyhydrides for large-area transmittance control. Solar Energy Materials and Solar Cells 2024;272. https://doi.org/10.1016/j.solmat.2024.112900.

[50] You CC, Mongstad T, Maehlen JP, Karazhanov S. Engineering of the band gap and optical properties of thin films of yttrium hydride. Appl Phys Lett 2014;105. https://doi.org/10.1063/1.4891175.

[51] Li SY, Niklasson GA, Granqvist CG. Thermochromic fenestration with VO 2-based materials: Three challenges and how they can be met. Thin Solid Films, vol. 520, 2012, p. 3823–8. https://doi.org/10.1016/j.tsf.2011.10.053.

[52] Nafezarefi F, Schreuders H, Dam B, Cornelius S. Photochromism of rare-earth metal-oxy-hydrides. Appl Phys Lett 2017;111. https://doi.org/10.1063/1.4995081.

[53] Sørby MH, Martinsen F, Karazhanov SZ, Hauback BC, Marstein ES. Article 2 On the crystal chemistry of photochromic yttrium oxyhydride 2022;15. https://doi.org/10.3390/xxxxx.

[54] Colombi G, Cornelius S, Longo A, Dam B. Structure Model for Anion-Disordered Photochromic Gadolinium Oxyhydride Thin Films. Journal of Physical Chemistry C 2020;124:13541–9. https://doi.org/10.1021/acs.jpcc.0c02410.

[55] Colombi G, Stigter R, Chaykina D, Banerjee S, Kentgens APM, Eijt SWH, et al. Energy, metastability, and optical properties of anion-disordered R Ox H3-2x (R= Y, La) oxyhydrides: A computational study. Phys Rev B 2022;105. https://doi.org/10.1103/PhysRevB.105.054208.

[56] You CC, Mongstad T, Marstein ES, Karazhanov SZ. The dependence of structural, electrical and optical properties on the composition of photochromic yttrium oxyhydride thin films. Materialia (Oxf) 2019;6. https://doi.org/10.1016/j.mtla.2019.100307.

[57] Tang M, Wang X, Hu X, Liu G, Liu Z, Deng H. Exploring the kinetic mechanism of dehydrogenation based on yttrium hydride surfaces: First principles calculations. Journal of Nuclear Materials 2025;614:155903. https://doi.org/10.1016/j.jnucmat.2025.155903.

[58] Gilmer GH, Huang H, de la Rubia TD, Dalla Torre J, Baumann F. Lattice Monte Carlo models of thin film deposition. Thin Solid Films 2000;365:189–200. https://doi.org/10.1016/S0040-6090(99)01057-3.

[59] Lundin D, Čada M, Hubička Z. Ionization of sputtered Ti, Al, and C coupled with plasma characterization in HiPIMS. Plasma Sources Sci Technol 2015;24. https://doi.org/10.1088/0963-0252/24/3/035018.

[60] Montero J, Martinsen FA, García-Tecedor M, Karazhanov SZ, Maestre D, Hauback B, et al. Photochromic mechanism in oxygen-containing yttrium hydride thin films: An optical perspective. Phys Rev B 2017;95. https://doi.org/10.1103/PhysRevB.95.201301.

[61] Baba EM, Montero J, Strugovshchikov E, Zayim EÖ, Karazhanov S. Light-induced breathing in photochromic yttrium oxyhydrides. Phys Rev Mater 2020;4:025201. https://doi.org/10.1103/PhysRevMaterials.4.025201.

[62] Chai J, Shao Z, Wang H, Ming C, Oh W, Ye T, et al. Ultrafast processes in photochromic material YHxOy studied by excited-state density functional theory simulation. Sci China Mater 2020;63:1579–87. https://doi.org/10.1007/s40843-020-1343-x.

[63] La M, Li N, Sha R, Bao S, Jin P. Excellent photochromic properties of an oxygen-containing yttrium hydride coated with tungsten oxide (YHx:O/WO3). Scr Mater 2018;142:36–40. https://doi.org/10.1016/j.scriptamat.2017.08.020.

[64] Colombi G, De Krom T, Chaykina D, Cornelius S, Eijt SWH, Dam B. Influence of Cation (RE = Sc, Y, Gd) and O/H Anion Ratio on the Photochromic Properties of REOxH3-2 xThin Films. ACS Photonics 2021;8:709–15. https://doi.org/10.1021/acsphotonics.0c01877.

[65] You CC, Mongstad T, Maehlen JP, Karazhanov S. Dynamic reactive sputtering of photochromic yttrium hydride thin films. Solar Energy Materials and Solar Cells 2015;143:623–6. https://doi.org/10.1016/j.solmat.2014.08.016.

[66] Kazi S, Moldarev D, Moro M V., Primetzhofer D, Wolff M. Correlating Photoconductivity and Optical Properties in Oxygen-Containing Yttrium Hydride Thin Films. Physica Status Solidi - Rapid Research Letters 2023;17. https://doi.org/10.1002/pssr.202200435.

[67]  You CC, Baba EM, Nordseth Ø, Karazhanov SZh. Photochromic Properties of Yttrium Oxyhydride Powders Derived from Magnetron-Sputtered Thin Films. Physica Status Solidi (RRL) – Rapid Research Letters 2025. https://doi.org/10.1002/pssr.202500206.

[68]  Nafezarefi F, Cornelius S, Nijskens J, Schreuders H, Dam B. Effect of the addition of zirconium on the photochromic properties of yttrium oxy-hydride. Solar Energy Materials and Solar Cells 2019;200. https://doi.org/10.1016/j.solmat.2019.109923.